\documentstyle[aps,twocolumn,floats,prl]{revtex}
\input epsf.sty

\newcommand{\pddm}{p_{\rm ddm}}
\newcommand{\rhoddm}{\rho_{\rm ddm}}
\newcommand{\pr}{p_{\rm r}}
\newcommand{\rhor}{\rho_{\rm r}}
\newcommand{\wgdm}{w_{\rm ddm}}
\newcommand{\cgdm}{c_{\rm ddm}}
\newcommand{\alphagdm}{\alpha_{\rm ddm}}

\begin{document}

\twocolumn[\hsize\textwidth\columnwidth\hsize\csname
@twocolumnfalse\endcsname

\title{An Isocurvature Mechanism for Structure Formation}

\author{Wayne Hu}
\address{Institute for Advanced Study, Princeton, NJ 08540}
\maketitle
\begin{abstract}
We examine a novel mechanism for structure formation 
involving initial number density fluctuations between 
relativistic species, one of which then undergoes a 
temporary downward variation in its equation
of state and generates superhorizon-scale 
density fluctuations. Isocurvature decaying dark matter 
models (iDDM) provide concrete examples.  This mechanism
solves the phenomenological problems of traditional 
isocurvature models, allowing iDDM models to fit the current
CMB and large-scale structure data, while still providing novel
behavior.  We characterize the decaying dark matter and its 
decay products as a single component of ``generalized dark matter''.
This simplifies calculations in decaying dark matter  
models and others that utilize this mechanism for structure formation.
\end{abstract}
\vskip 0.5truecm
]
%\keywords{cosmology:theory --- dark matter ---
%large-scale structure of the universe}

Conventional models for structure formation utilize either
initial density fluctuations that then grow by gravitational
instability or initial stress fluctuations between the matter
and radiation which push the matter
into gravitationally unstable configurations.  We call these
adiabatic and traditional isocurvature models, respectively.
The possibility remains that the origin of structure
lies elsewhere.  We consider here an origin from 
a temporary change in the equation of state of
the dark matter and give a concrete example in the form
of an isocurvature decaying dark matter 
(iDDM) model for structure formation.

{\it Traditional Isocurvature Models.---} In traditional isocurvature 
models, the initial density balance 
comes at the expense of number density or ``entropy'' 
fluctuations between the matter and radiation.
These models suffer from various problems that stem from 
the lack of a natural timescale besides the expansion time in 
the radiation-dominated early universe.   Phenomenologically,
models with scale-invariant entropy fluctuations 
differ from their adiabatic counterparts with scale-invariant
curvature fluctuations through: (1)  
an under-production of large-scale 
structure relative to large-angle CMB anisotropies \cite{EfsBon86} 
and (2) diminished acoustic peaks appearing at small angles
\cite{HuSug95}.  
The former is in conflict with the observed large-scale structure,
once normalized to large-scale anisotropies, and
the latter is in conflict with recent measurements of 
degree-scale anisotropies. 

Variants of the basic isocurvature model have been proposed to
alleviate these problems.  Peebles \cite{Pee87}
introduced a blue-tilt to the
entropy power spectrum to address the lack of large-scale structure
and the falling spectrum of CMB anisotropies.  Unfortunately,
such models are in moderate conflict with the slope of the COBE
anisotropy spectrum \cite{HuBunSug95}.  The 
basic problem with the spectrum
is one of timescale.  This model forms structure through the residual
stress perturbation remaining when the density fluctuations in
a fluid with relativistic pressure $p_{\rm r}=\rhor/3$ are balanced off
those in a pressureless fluid.  These stresses move matter around
until the perturbation crosses the horizon or the radiation density
becomes negligible.  For wavelengths smaller than the horizon at
matter-radiation equality, the amount of time the stresses have
to act is a decreasing function of wavenumber $k$, leading to a falling 
spectrum of CMB anisotropies and large-scale structure.  

Cosmological defects provide an alternative means of generating
structure from isocurvature initial conditions that reduce
these problems of spectral shape. 
These models balance seed fluctuations in the defects
off ordinary matter to establish isocurvature initial conditions. 
Unlike the matter-radiation isocurvature models, the temporal
behavior of the stresses scale with the horizon crossing time 
\cite{Kib85}.  
Thus the effect of the stresses on the ordinary matter is 
the same for all scales.  The other problems of isocurvature
models remain.  For example, these models tend to 
underpredict large-scale structure (see \cite{PenSelTur97,AlbBatRob97a} for recent assessments). 
Part of this problem would seem to be common to all 
isocurvature models.  It comes from the relationship between
the Sachs-Wolfe temperature anisotropy $\Delta T/T$
and the Newtonian curvature $\Phi$, $\Delta T/T \approx -2\Phi$,
which should be compared with
$\Delta T/T \approx \Phi/3$ in adiabatic models.
Defect models fare even worse since their vector and tensor
modes provide additional sources of anisotropy.  
Furthermore, the superhorizon growth of the curvature leads to 
forced acoustic oscillations whose features are shifted to smaller
angles relative to adiabatic models.
This would also seem to be 
a generic feature of isocurvature models since the 
curvature can only grow from its vanishing initial condition. 
The observed rise of the anisotropy power spectrum
at degree scales would be difficult to explain in defect or other 
similar isocurvature variants \cite{PenSelTur97}.  

{\it Novel Isocurvature Mechanism.---} Nonetheless, 
these problems are not fundamental to isocurvature
initial conditions but rather to the choice of the stress perturbation
history that generates the structure.  In both examples above,
the problem is that horizon crossing for the perturbations sets
the timescale over which the stresses act. 
If the stresses on all superhorizon scales could be turned on
and then off uniformly, then the universe would be left
with constant scale-invariant curvature fluctuations similar
to an adiabatic model.  

\begin{figure}
\centerline{ \epsfxsize = 3.4truein \epsffile{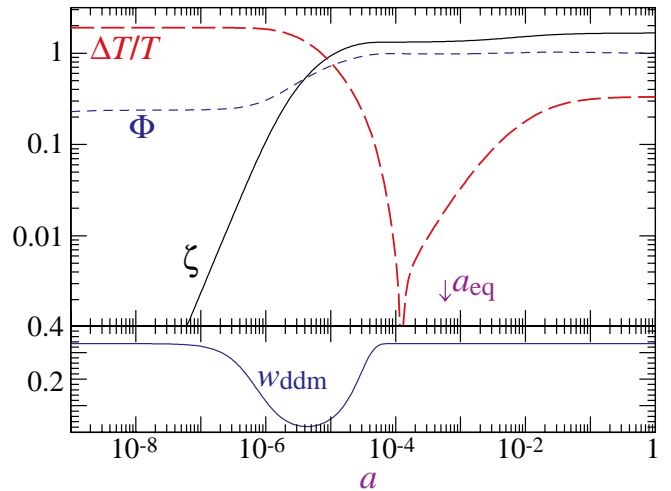}}
\caption{Time evolution of
the comoving curvature $\zeta$, Newtonian curvature $\Phi$, 
effective temperature perturbation $\Delta T/T$, and equation of
state $\wgdm$ in $m=5$ keV, $\tau=5$yrs, $\Omega_{\rm m}=1$, $h=0.5$,
$\Omega_bh^2=0.02$ universe.}
\label{fig:timefig}
\end{figure}

A dark matter species that undergoes a variation in its 
equation of state provides such a mechanism.
Consider the case where a dark matter species goes non-relativistic and then
decays: the equation of state for this matter and its decay products begins
at $\wgdm=\pddm/\rhoddm=1/3$, 
dips toward zero, and returns to $1/3$ (see Fig.~\ref{fig:timefig}
lower panel).
Another important aspect of this model is that since the perturbations
are in a species that is originally ultra-relativistic, isocurvature
conditions require balancing perturbations from
the other relativistic species, including
the photons in the CMB.
Because 
the balance is through species with the same equation of state
initially, vanishing density perturbations imply vanishing
stress perturbations as well.
The initial stress-energy tensor of the total matter is 
completely homogeneous and isotropic.

As the dark matter becomes non-relativistic, the initially
counterbalancing stress fluctuations become unbalanced as
$\delta \pddm$ drops below $\delta\rhoddm/3$. 
They move matter around and form density or curvature fluctuations.
When the dark matter then decays back into radiation, the 
stresses regain their balance and stop forming curvature fluctuations.
The process that generates curvature in
this model in fact has an exact solution \cite{HuEis98}
in the simple case that the fluctuations in the decaying dark
matter are balanced off a single radiation component (r).
The curvature in the comoving gauge 
$\zeta$ is directly related to the
stresses \cite{Bar80} and is given by
\begin{equation}
\zeta(a) = {\sigma \over 3} { \pddm +\rhoddm  \over p+\rho}\Big|^a_0
	  + \zeta(0)\,, 
\end{equation}  
where $a$ is the scale factor and
\begin{equation}
\sigma = \left({\delta \rhoddm \over \rhoddm + \pddm} -
	 {\delta \rhor \over \rhor + \pr} \right) \Bigg|_{0}\, .
\end{equation}
Isocurvature models have $\zeta(0)=0$ by definition.
The generation of curvature is most effective if the DDM species
comes to dominate the total energy density before the decay.  
The curvature remains constant after the decay in all cases
since the decay products and ordinary radiation
redshift in the same way leaving $\rhoddm/\rho$ constant 
until other matter species become important.

This mechanism solves the problems isocurvature models have with
the acoustic peaks in the CMB.   The constancy of the potential
outside the horizon during radiation domination insures acoustic
phenomenology that is similar to the adiabatic model: scale-invariant 
oscillations before diffusion damping and a cosine
series of harmonic peaks.  As noted by many authors in the
adiabatic variant of the DDM scenario 
(e.g.~\cite{BonEfs91,WhiGelSil95}), the 
decay into relativistic species also delays 
matter-radiation equality and allows a high-$\Omega_{\rm m}$ model
to look like a low-$\Omega_{\rm m}$ model with respect to the
large-scale structure power spectrum and CMB anisotropies.

Unfortunately, this mechanism does not automatically solve the other
problem of isocurvature models: the ratio of large-scale
anisotropies to large-scale structure.   In fact, it can
actually make the agreement worse.  If the DDM fluctuations are
balanced by CMB fluctuations initially, then the 
CMB temperature will be lower in regions where the DDM density
is higher.  Since these are the sites of potential wells
when the DDM become non-relativistic, those cold photons will
lose more energy climbing out of the potential wells after last
scattering and become colder.  
The complete gravitational redshift effect is
\begin{equation}
{\delta T \over T}(1) \approx -2\Phi \Big|^{1}_0 + {\delta T \over T}(0) 
	\,.
\end{equation}
The solution is clear.  The initial conditions must 
involve the photons and DDM together compensating the other species of
radiation.  The photons will then be initially hotter at the
sites that form potential wells. 

{\it iDDM Models.---} For definiteness, let us now 
examine the phenomenology of the iDDM class of models.  
The important aspect is that
the DDM scenario allows the required variation in the equation
of state. To highlight this property, we model the decaying 
particle and its decay products 
as a single component of ``generalized''
dark matter of the type introduced in \cite{Hu98}; we label
it DDM here.  
The DDM is described by an equation of state $\wgdm$, a sound
speed $\cgdm$, and a viscosity parameter $\alphagdm$.
The equation of state of the DDM is given implicitly in terms
of the mass $m$ and lifetime $\tau$ of the particle,
\begin{eqnarray}
\wgdm & = & { 3(1+w_1)\rho_1 - 4\rho_2 \over \rho_1 + \rho_2}\,, 
	\nonumber\\
{d \rho_1 \over d\ln a} &=& -3(1+w_1)\rho_1 - {1 \over H\tau} \rho_1 
			\,,\\
{d \rho_2 \over d\ln a} &=& -4\rho_2 + {1 \over H\tau} \rho_1 \,,
	\nonumber
\end{eqnarray}
where  $H = a^{-1}(d a / dt )$, 
\begin{equation}
w_1 =  {1 \over 3} [ 1 + (a /a_{\rm nr})^{2p}]^{-1/p},
\end{equation}
with $p=0.872$ and $a_{\rm nr} = 8.3 \times 10^{-7} 
	(m/{\rm keV})^{-1} (T/T_\gamma)$.  Here 
$T/T_\gamma$ is the 
ratio of temperatures of the DDM and photons while the DDM
was relativistic.  We assume here that the DDM accounts for
one of the usual neutrino species so that 
$T/T_\gamma = (4/11)^{1/3}$.  The sound speed is given
by $\cgdm^2 = \wgdm - (d \wgdm / d\ln a)/(3+3\wgdm)$ and the
viscosity parameter $\alphagdm=\wgdm$ following \cite{Hu98}.
We have checked this approximation against explicit
calculations of an adiabatic DDM model with the techniques of
\cite{BonEfs91}.
The approximation has the
practical benefit of being simple to implement (c.f. \cite{BhaSet98}) 
and the pedagogical value of highlighting the important
aspects of this mechanism. 

The critical parameter is $m^2 \tau$ \cite{WhiGelSil95},
since it determines
the fractional increase in the total radiation density due to the
decay $\rhor \rightarrow \rhor (1+0.15 m_{\rm keV}^{4/3} 
	\tau_{\rm yr}^{2/3}$); as we have seen
changes in the energy density of the DDM relative to the other
components is what drives curvature generation.

We establish isocurvature initial conditions of the form 
\begin{eqnarray}
{\delta \rhoddm \over \rhoddm} &=& {\delta \rho_\gamma \over \rho_\gamma}
= {3 \over 4} {\delta \rho_{\rm m} \over \rho_{\rm m}}\,, \nonumber\\
\delta \rho_\nu & = & -(\delta \rhoddm + \delta\rhoddm + 
	\delta\rho_{\rm m}) \,,
\label{eqn:initialconditions}
\end{eqnarray}
where $\rho_\nu$ is the density in the remaining two neutrino
species and $\rho_{\rm m}$ is the density in baryons and CDM with all
other metric and matter perturbations zero.  
More complicated initial conditions can alter the ratio between
$\Delta T/T$ and $\Phi$.
Despite these simple
initial conditions, some care must be taken to insure numerical
stability in the evolution.  In the synchronous gauge, we 
recommend the use of the variables $H_T$ and $-H_L-H_T/3$ in
the notation of \cite{Bar80}.

The time evolution of superhorizon scale perturbations 
in an example model 
with $m=5$keV and $\tau=5$ yrs is shown in Fig.~\ref{fig:timefig} 
(upper panel).  As discussed above, the
comoving curvature $\zeta$ grows rapidly as the equation of state $\wgdm$ dips
below its relativistic value.  On the other hand, 
the curvature
in the Newtonian gauge $\Phi$ is finite and constant {\it before}
the decay.  
In traditional isocurvature 
and adiabatic models, the $\zeta$ and $\Phi$ 
are simply proportional.  The difference here
is that the radiation possesses substantial density perturbations which leads
to anisotropic stress perturbations in the neutrinos
of order $\pi = (k\eta)^2 {\cal O}(\delta \rhoddm/\rhoddm)$ where $\eta$ is the
conformal time; balancing anisotropies in the photons are prevented
by Compton scattering. 
The relativistic Poisson equation involves curvature sources of order 
$\pi/(k\eta)^2$ leading to constant Newtonian curvature perturbations.  However
once the dark matter decays, the Newtonian and comoving curvatures again become
proportional to each other. 

The second interesting feature is that, unlike adiabatic models,
the comoving curvature $\zeta$ grows 
at the second transition to matter domination $a_{\rm eq}$.  The initial 
conditions (\ref{eqn:initialconditions}) implies an entropy fluctuation
between the non-relativistic matter $\rho_{\rm m}$ and the combined radiation after
the decay.  This entropy fluctuation causes stress fluctuations when the
non-relativistic matter comes to dominate the universe.  This behavior
is identical to matter-radiation isocurvature models and implies
that the phenomenology of the complete model will be a combination of traditional
isocurvature and adiabatic models. 
Finally, as desired, the initial temperature perturbation 
$\Delta T/T$ cancels part of the redshift effect from the growing Newtonian potential
leaving smaller CMB anisotropies for a given matter density fluctuation.  

\begin{figure}
\centerline{ \epsfxsize = 3.5truein \epsffile{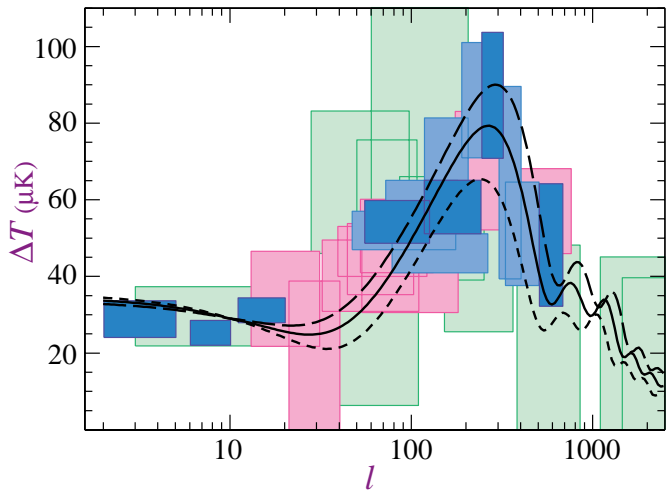}}
\caption{CMB anisotropies for models $m=5$keV and $\tau=$3,5,8 yrs
(short-dashed, solid, long-dashed lines) in an $\Omega_{\rm m}=1$, $h=0.5$
and $\Omega_b h^2=0.02$ universe compared with the current CMB
data \protect\cite{webpage}
with 1$\sigma$ error boxes shaded according to area.}
\label{fig:cl}
\end{figure}

The CMB anisotropies of models with $m=5$keV and $\tau=$3, 5, 8 yrs are shown 
in Fig.~\ref{fig:cl} and compared with the current data.  
Models with the same $m^2 \tau$ are approximately degenerate.
The large angle anisotropies
have the spectral shape of traditional isocurvature models with a decline
toward smaller angles due to the matter-radiation transition.   However, 
the acoustic peaks show a distinctly adiabatic pattern with peak positions in
a classic $1 : 2 : 3 \ldots$ series and high odd peaks.  The result is a dip
in the anisotropies at intermediate scales that is an interesting signature
of such models.  The CMB polarization however 
mimics adiabatic models predictions.

\begin{figure}
\vskip -0.15truecm
\centerline{ \epsfxsize = 3.5truein \epsffile{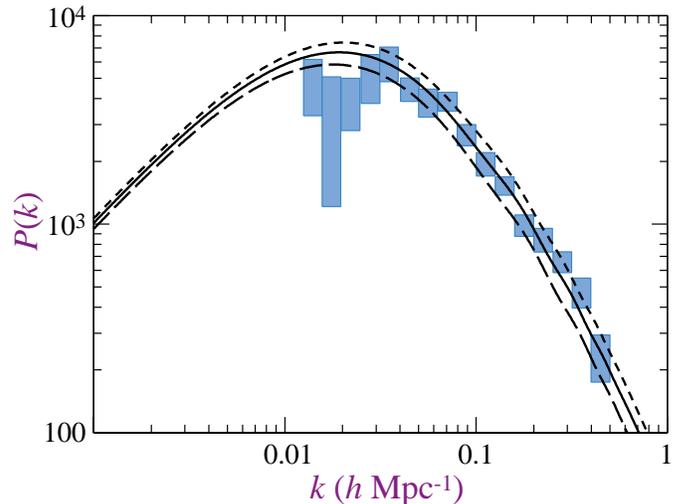}}
\caption{Matter power spectrum for the same models as 
Fig.~\protect\ref{fig:cl} compared with galaxy survey data from 
\protect\cite{PeaDod94} with  1$\sigma$ error boxes.  
Note that the data may be shifted up or down
to account for the unknown galaxy bias. }
\label{fig:power}
\end{figure}

The power spectrum of the large scale structure in the same models are given
in Fig.~\ref{fig:power} and compared with the data \cite{PeaDod94}.  
The shape of the power spectrum matches the data adequately; for
example, the ratio of power at the $50 h^{-1}$ Mpc to $8 h^{-1}$ Mpc scale
$\sigma_{50}/\sigma_{8} = 0.16$ in the $\tau=3$yrs model. 
Relative to an adiabatic model with the same parameters, these
have more large compared with small scale power due to growth
at the second matter-radiation transition.
Note that the observations may be slid up and down to account for an unknown
galaxy bias.   On the other hand, the model curves are COBE-normalized through
the fitting form of \cite{BunWhi97} and predict $\sigma_8 =0.62$ for the
$3$yrs model.  This value would reproduce the present-day
cluster abundance adequately (e.g. \cite{EkeColFre96}) and at
 high-redshift marginally
(e.g. \cite{BahFanCen97}).   The high value of $\Omega_{\rm m}$ in these examples
is observationally disfavored by recent determinations of the the
luminosity distance to high-redshift supernovae 
\cite{Rieetal98,Peretal98};
if these preliminary indications are borne out by future measures,
lower $\Omega_{\rm m}$ variants of these models can be considered.

{\it Discussion.---}
Perhaps more interesting than the details of any specific model of
this type is the lesson
their mere existance teaches us.   That the current CMB and large-scale structure data
is consistent with a high-density model with scale-invariant
initial isocurvature perturbations warns us against overinterpreting the current
data.  It also suggests that if future observations reveal phenomenology that
is close to but not precisely predicted by standard adiabatic models, we should
not necessarily abandon the search for new paradigms for structure formation.  
Our example shows that the acoustic peak locations do {\it not} actually discriminate 
between adiabatic and isocurvature initial conditions in general.  They do however
relate to the role of causality in the generation of perturbations
as discussed in \cite{HuWhi96}.  
To be explained causally, the
scale-invariant isocurvature ``initial'' 
conditions employed here still require
a period of superluminal expansion, like that provided by inflation. 
The models considered in the last section require that fluctuations 
in the chemical potential of say the $e$ and $\mu$ neutrinos balance
temperature fluctuations and may be difficult to arrange in inflationary
models.  However the general mechanism is of wider interest.
Exploration of alternatives such as these helps to sharpen
the questions that can be asked of the data. 

In summary, the proposed isocurvature mechanism for structure formation 
solves the problems with the spectra and relative amplitude
of the matter and CMB fluctuations associated with traditional
isocurvature models. 
The key aspects of this mechanism are that (1) joint fluctuations in
the energy density of some initially relativistic component 
{\it and} the photons are
balanced by the remaining radiation and (2) the equation of state of
the DDM drops and then returns to the ultrarelativistic value
of $\wgdm=1/3$ before the scales relevant to large-scale structure
and CMB anisotropies enter the horizon.  We have shown that
this situation can be realized with a dark matter particle in 
the keV mass range that decays on the timescale of a year. 
Such models are in agreement with the current data but have novel
features that are testable with the upcoming generation of
experiments.  Their existance calls into question widely-held beliefs
about isocurvature models and cautions agains overinterpretation
of current and future data sets. 

{\it Acknowledgments:} I thank D.J. Eisenstein \& M. White for
useful discussions.  This work was supported by the W.M. Keck Foundation,
NSF PHY-9513835 and a Sloan Foundation Fellowship.


\begin{thebibliography}{99}


\bibitem{EfsBon86}
            G. Efstathiou \& J.R. Bond, Mon. Not. Roy. Astr. Soc.,
            218, 103 (1986)
\bibitem{HuSug95} W. Hu \& N. Sugiyama, Phys. Rev. D. 51, 2599 (1995)
%                [astro-ph/9411008]
\bibitem{Pee87} P.J.E. Peebles, Nature, 327, 210 (1987)
\bibitem{HuBunSug95} W. Hu, E. Bunn, \& N. Sugiyama,
                 Astrophys. J. Lett. 447, L59 (1995)
\bibitem{Kib85} T.W.B. Kibble, Nucl. Phys. B262, 227 (1985)
\bibitem{PenSelTur97} U.-L. Pen, U. Seljak, \& N. Turok,
                Phys. Rev. Lett. 79, 1611 (1997)
\bibitem{AlbBatRob97a} A. Albrecht, R.A. Battye, \& J. Robinson
         Phys. Rev. Lett. 79, 4736 (1997)
\bibitem{HuEis98} W. Hu \& D.J. Eisenstein (in preparation)
\bibitem{BonEfs91} J.R. Bond \& G. Efstathiou, Phys. Let. B265, 245
                (1991)
\bibitem{WhiGelSil95} M. White, G.Gelmini, \& J.Silk, 
		Phys. Rev. D., 51, 2669 (1995)
\bibitem{Bar80} J.M. Bardeen, Phys. Rev. D., 22, 1882 (1980)
\bibitem{Hu98} W. Hu, Astrophys. J. (in press)
               [astro-ph/9801234]
\bibitem{BhaSet98} S. Bharadwaj \& S.K. Sethi, Astrophys. J. Supp. 114, 37  (1998)
\bibitem{PeaDod94} J.A. Peacock \& S.J. Dodds, Mon. Not. Roy. Astron. Soc., 267, 1020 (1994)
\bibitem{HuWhi96} W. Hu \& M. White, Astrophys. J., 471, 30 (1996); ibid,
		Phys. Rev. Lett., 77, 1687 (1996)
\bibitem{BunWhi97} E.F. Bunn \& M. White, Astrophys. J., 480, 6 (1997)
%[astro-ph/9607060]
\bibitem{webpage} see e.g.~{\tt http://www.sns.ias.edu/$\sim$whu}
\bibitem{EkeColFre96} V.R. Eke, S. Cole, \& C.S. Frenk, Mon. Not. Roy. Astron. Soc., 282, 263 (1996)
\bibitem{BahFanCen97} N.A. Bahcall, X. Fan, \& R. Cen, Astrophys. J., 485, L53 (1997)
\bibitem{Rieetal98} A.G. Riess et al.,  Astronom. J., in press [astro-ph/9805201]
\bibitem{Peretal98} S. Perlmutter, et al., Nature, 391, 51 (1998)

\end{thebibliography}
\end{document}